\documentclass[sigconf]{acmart}
\usepackage{url}
\usepackage{bm}
\usepackage{color}
\usepackage{booktabs}
\usepackage{multirow}
\usepackage{subfigure}
\usepackage{enumitem}
\usepackage[ruled,vlined]{algorithm2e} %
\usepackage[switch]{lineno} 

\newcommand{\red}[1]{#1}
\newcommand{\blue}[1]{#1}

\AtBeginDocument{%
  \providecommand\BibTeX{{%
    \normalfont B\kern-0.5em{\scshape i\kern-0.25em b}\kern-0.8em\TeX}}}

\copyrightyear{2022}
\acmYear{2022}
\setcopyright{acmcopyright}\acmConference[WSDM '22]{Proceedings of the Fifteenth ACM International Conference on Web Search and Data Mining}{February 21--25, 2022}{Tempe, AZ, USA}
\acmBooktitle{Proceedings of the Fifteenth ACM International Conference on Web Search and Data Mining (WSDM '22), February 21--25, 2022, Tempe, AZ, USA}
\acmPrice{15.00}
\acmDOI{10.1145/3488560.3498388}
\acmISBN{978-1-4503-9132-0/22/02}

\begin{document}
\fancyhead{}

\title{RecGURU: Adversarial Learning of Generalized User Representations for Cross-Domain Recommendation}

\author{Chenglin Li}
\affiliation{
  \institution{University of Alberta}
  \city{Edmonton, AB}
  \country{Canada}
}

\author{Mingjun Zhao}
\affiliation{
  \institution{University of Alberta}
  \city{Edmonton, AB}
  \country{Canada}
}

\author{Huanming Zhang}
\affiliation{
  \institution{Platform and Content Group, Tencent}
  \city{Shenzhen}
  \country{China}
}

\author{Chenyun Yu}
\affiliation{
  \institution{Sun Yat-sen University}
  \city{Shenzhen}
  \country{China}
}

\author{Lei Cheng}
\affiliation{%
  \institution{Platform and Content Group, Tencent}
  \city{Shenzhen}
  \country{China}
}

\author{Guoqiang Shu}
\affiliation{%
  \institution{Platform and Content Group, Tencent}
  \city{Shenzhen}
  \country{China}
}

\author{Beibei Kong}
\affiliation{
  \institution{Platform and Content Group, Tencent}
  \city{Shenzhen}
  \country{China}
}

\author{Di Niu}
\affiliation{
  \institution{University of Alberta}
  \city{Edmonton, AB}
  \country{Canada}
}
\renewcommand{\shortauthors}{Chenglin Li, et al.}

\begin{abstract}
Cross-domain recommendation can help alleviate the data sparsity issue in traditional sequential recommender systems. In this paper, we propose the RecGURU algorithm framework to generate a Generalized User Representation (GUR) incorporating user information across domains in sequential recommendation, even when there is minimum or no common users in the two domains. We propose a self-attentive autoencoder to derive latent user representations, and a domain discriminator, which aims to predict the origin domain of a generated latent representation. We propose a novel adversarial learning method to train the two modules to unify user embeddings generated from different domains into a single global GUR for each user. The learned GUR captures the overall preferences and characteristics of a user and thus can be used to augment the behavior data and improve recommendations in any single domain in which the user is involved. Extensive experiments have been conducted on two public cross-domain recommendation datasets as well as a large dataset collected from real-world applications. The results demonstrate that RecGURU boosts performance and outperforms various state-of-the-art sequential recommendation and cross-domain recommendation methods. The collected data will be released to facilitate future research. 
\end{abstract}



\maketitle

\section{Introduction}
Sequential recommendation has achieved great success in modeling the preferences and intentions of online users by utilizing both their recent actions and long-term behavior history. Existing methods often leverage recurrent neural networks (RNNs) and attention mechanisms to model the interests of users based on sequential data \cite{sun2019bert4rec,ying2018sequential,hidasi2018recurrent,chen2018sequential}. However, most studies to date on sequential recommendations have been focusing on a single domain, where data sparsity issues prevail---chances are that most users may only have a short behavior history in the domain of interest.

Cross-domain recommendation has been proposed to alleviate the data sparsity issue by leveraging user behavior in multiple domains to help recommendation in the target domain and has attracted much attention in both academia and industry.
Recent work on cross-domain recommendation focuses on the transfer learning of user and item information from diverse perspectives. 
For example, mapping functions have been proposed to map user representations from one domain to another, by learning from the behavior of users that appear in both domains \cite{man2017cross}.

Nevertheless, a common limitation of existing cross-domain recommendation methods is that they perform transfer learning primarily based on data of the overlapped users, and fail to function well when there are few or even no overlapped users in two domains~\cite{man2017cross}. However, in many real-world applications, there is often not a sufficient number of overlapped users. 

In this paper, we propose RecGURU, which consists of two parts: the Generalized User Representation Unit (GURU) to obtain a single Generalized User Representation (GUR) for each user, and the Cross-Domain Sequential Recommendation (CDSRec) unit to achieve cross-domain collaboration in sequential recommendation task. 
Instead of mapping user embeddings from one domain to another, we propose to generalize a user's embedding, i.e., its GUR, to incorporate information from both domains through an adversarial learning framework. 
Once a GUR is obtained for each user, we integrate the extracted GUR into the CDSRec unit using attention mechanisms to boost recommendation performance. Specifically, we make the following contributions:

First, in the GURU module, an autoencoder is proposed to generate informative user embeddings in each domain, which are to be unified later into a generalized embedding, the GUR, with adversarial learning. The autoencoder consists of a self-attentive encoder with model weights shared across both the source and target domains to produce latent user embeddings and two decoders to reconstruct behavior sequences of users in the source and target domains, respectively. 
We train the autoencoder through behavior sequence reconstructions to generate meaningful preliminary embeddings for users with unsupervised self-learning.

The GURU module further performs adversarial training to unify domain-dependent user embeddings into a single global (GUR) for each user, which domain-independent. 
Specifically, the encoder part of the proposed autoencoder serves as a generator to produce user embeddings, while a discriminator is trained to identify the origin domain of a generated embedding for a user randomly sampled from either the source or the target domain. The encoder and discriminator are trained alternately with adversarial objectives until the discriminator can not distinguish which domain a given user embedding comes from. This is when the user embeddings in the two domains become statistically indistinguishable and GURs are supposed to be generalized global user embeddings, incorporating information from both domains. 
This method does not rely on common users present in both target and source domains, therefore eliminating the dependency on overlapped users as required by prior art.

Furthermore, we introduce an effective and stable training procedure for RecGURU, consisting of three phases. We first pre-train the autoencoder to substantially reduce the reconstruction loss which boost-starts the subsequent adversarial learning. In the adversarial learning phase, the reconstruction loss is further jointly optimized in a multi-task fashion which prevents the encoder from generating wild representations, stabilizing adversarial learning. In the meantime, RecGURU can still leverage overlapped users as prior work does, by introducing an $l_2$ penalty in the optimization procedure to explicitly force each common user to have the same shared embedding in different domains. Finally, the CDSRec module, \blue{which incorporates the GUR with attention mechanisms,} is fine-tuned in the target domain with the next-item recommendation task to boost the recommendation performance.

Through extensive experiments, we show that RecGURU has achieved improvement on sequential recommendation, compared to several state-of-the-art single-domain and cross-domain recommendation methods. 
Specifically, we outperform all the baselines on various metrics by a large margin on the Amazon datasets including "Sport", "Clothing", "Movie", and "Book".
Additionally, we have collected a large cross-domain recommendation dataset with two domains, i.e. ``Wesee'' and ``Tencent Video'', from two real-world applications which provide video streams to millions of users.
Ablation studies on the collected datasets with various portions of overlapped users are conducted to show the effectiveness of each proposed sub-module as well as the robustness of our method. 
The collected datasets will be made public to facilitate future research in the field of cross-domain and sequential recommendation.

\section{Related work}
\label{sec:related}

\textbf{Sequential Recommendation}. 
Early studies adopt Markov Chains (MCs) to capture sequential patterns from users’ historical interactions \cite{rendle2010factorizing,he2017translation}. 
Recently, researchers have been putting effort to adapt the Recurrent Neural networks (RNN) and attention mechanism to solve sequential recommendation problems. 
GRU4Rec \cite{hidasi2015session} and its improved version GRU4Rec+ \cite{hidasi2018recurrent} leverage Gated Recurrent Unit (GRU) with BPR loss to model a user's sequential behaviors. 
In SAS \cite{kang2018self}, unidirectional self-attention is adopted here to encode a user's historical behavior. 
Bert4Rec \cite{sun2019bert4rec} follows the idea of BERT and trained a bidirectional self-attention model. 
SHAN \cite{ying2018sequential} adopts a hierarchical attention framework where two attention networks are used to model user's long- and short-term preferences. 

\textbf{Cross-domain Recommendation}. 
Cross-domain recommendation alleviates data sparsity issues posed in single domain recommendation via auxiliary information from other domains.
CoNet \cite{hu2018conet} transfers and combines knowledge across different domains through cross-connections between feed-forward neural networks. 
However, these methods only focus on overlapped users. 
CATN \cite{zhao2020catn} solves the cold-start problem via aspect transfer which requires side information of both users and items. 
\cite{krishnan2020transfer} leverages meta-transfer learning to address the sparsity problem, but it requires multiple contextual information such as the user's average spend. 

To transfer knowledge from the source domain to the target domain, 
EMCDR \cite{man2017cross} learns a mapping function on overlapped users which maps user preferences across domains. 
DCDCSR \cite{zhu2018deep} maps the latent factors in the target domain to fit the benchmark factors which combines the features in both the target and source domains. 
\red{To reduce the dependency on overlapped users, SSCDR \cite{kang2019semi} adopts a semi-supervised strategy. }
\red{DDTCDR \cite{li2020ddtcdr} and its improved version DOML\cite{li2021dual} adopt dual metric learning (DML).}

Generative Adversarial Network (GAN)~\cite{goodfellow2014generative} is gaining popularity in cross-domain recommendation\cite{perera2019cngan, wang2019recsys,yan2020cross,li2020atlrec,wang2019recsys}. 
CnGAN~\cite{perera2019cngan} \red{introduces the use of the GAN to learn a better mapping function of user representations from the source domain to the target domain. 
Since the discriminator in CnGAN is trained to distinguish between real and synthetically mapped pairs, where the real mapped pairs only come from overlapped users. Thus, CnGAN is critically dependent on the quantity and quality of overlapped users, which usually cannot be guaranteed in reality. Additionally, the CnGAN tries to stabilize the training of GAN by introducing more synthetically mapped pairs. However, this makes the input data to the discriminator unbalanced, thus may fail to train a good mapping function.
target domain. 
\citeauthor{yan2020cross} \cite{yan2020cross} adopt adversarial samples in the training process to improve the generalization ability of the cross-domain recommender system.  
ATLRec~\cite{li2020atlrec} transfers shareable features across domains, however, it focuses on overlapped users which show up in both the target and source domains. }

\textbf{Cross-domain Sequential Recommendation}. 
$\pi$-Net \cite{ma2019pi} is able to generate recommendations for both domains through a cross-domain transfer unit. However, it requires synchronously shared timestamps and can not be applied to non-overlapped users.
\citeauthor{zhuang2020sequential}~\cite{zhuang2020sequential} transfers users' novelty-seeking properties learned from the sequential data in the source domain to the target domain. 
\citeauthor{yuan2020parameter}~\cite{yuan2020parameter} proposed a framework that is able to fine-tune a large pre-trained user embedding network to adapt to downstream tasks in the target domain. 
\red{However, these studies have different problem settings from our work. And they can only be applied to overlapped users, but our method can handle both overlapped and non-overlapped users. }

\section{Method}
\label{sec:sys}

\begin{figure*}[ht]
  \centering
  \includegraphics[width=5.1in]{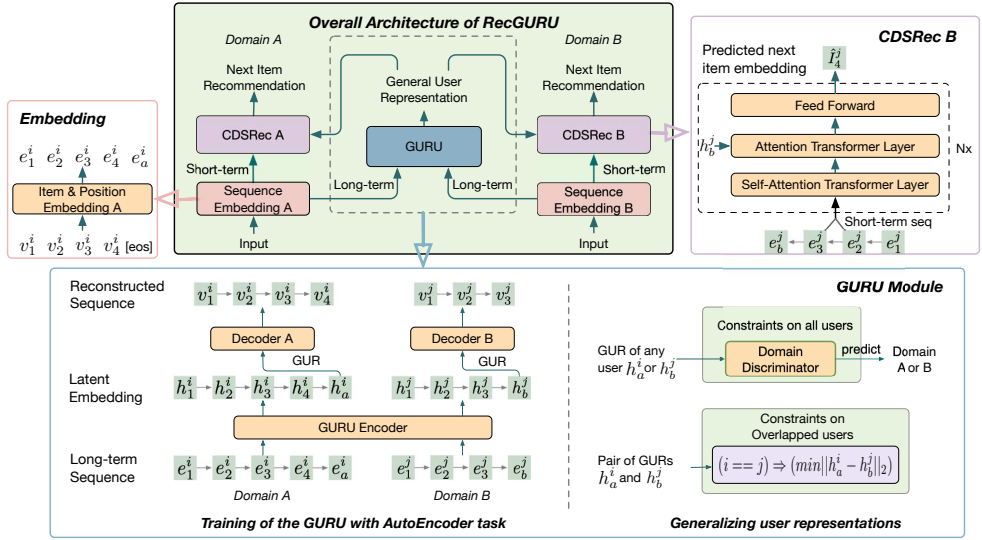}   
  \caption{Model structure of the proposed RecGURU. User behavior sequences in both domain $A$ and $B$ are fed into the GURU encoder to generate \red{generalized} latent user representations, i.e. $\bm{h}_a^i$ and $\bm{h}_b^i$.  
  Then, the generated GUR is fed into the CDSRec model for the next-item recommendation in each individual domain. 
   }
  \Description{A figure shows the overall structure of the General User Representation module.}
  \label{fig:sys}
\end{figure*}

\subsection{Problem Definition}
We first formulate the implicit feedback-based sequential recommendation problem in a single domain. 
Let $U = \{u_1,\cdots, u_{|U|}\}$ and $V = \{v_1, \cdots, v_{|V|}\}$ denote the sets of users and items, respectively, where $|U|$ and $|V|$ are the total number of users and items. 
For each user $u_i, \forall i \in \lbrace 1 ,\cdots ,|U|\rbrace$, its interactions with items, in chronological order, are denoted as $s^i = (v_1^i, \cdots, v_{|s^i|}^i)$, 
where $|s^i|$ is the length of behavior sequence $s^i$. 
Formally speaking, given a user $u_i$, sequential recommendation aims to predict the next item that the user is most likely to interact with at the next time step $|s^i| + 1$ based on his or her past behavior sequence $s^i$, which can be formalized as modeling the probability over all items:
\begin{equation} \label{eq:single_seq}
p(v_{|s^i| + 1}^i = v|s^i).
\end{equation}

In cross-domain scenarios, in order to improve the recommendation performance, user information from other domains is also taken into account. 
Specifically, for two domains $A$ and $B$, given an overlapped user who appears in both domains, $u_i \in (U_A \cap U_B)$, cross-domain sequential recommendation tries to improve the recommendation accuracy of the next item in one domain by integrating user information from both domains. 
For example, for next-item recommendation in domain $A$, it can be formulated as modeling the probability over all possible candidates given the behavior sequences $s_a^i$ and $s_b^i$ in domains $A$ and $B$:
\begin{equation} \label{eq:cross_seq}
p(v_{|s_A^i| + 1}^i = v|s_A^i, s_B^i). 
\end{equation}
For non-overlapped users, due to a lack of their behavior in both domains, we can only exploit the implicit information from the other domain to enhance the recommendation performance in the target domain. Therefore, Equation~\eqref{eq:cross_seq} is reformulated as
\begin{equation} \label{eq:cross_seq_non}
p(v_{|s_A^i| + 1}^i = v|s_A^i, \textit{info}(S_B)),
\end{equation}
where $S_B =\{s_B^j\}, \forall u_j \in U_B$ denotes the collection of behavior sequences for all users in domain $B$ and $\textit{info}(\cdot)$ represent a model which is used to extract any kinds of useful information from $S_B$ to assist recommendation in domain $A$.

\subsection{Overview of Proposed Method}  
Inspired by previous studies in single domain sequential recommendations which try to combine the long-term (static) and short-term (dynamic) preferences of users for next-item recommendation \cite{wang2015learning,xu2019survey,ying2018sequential}, 
we propose a novel RecGURU framework. It consists of two parts: a generalized user representation unit (GURU) which learns the generalized user representation across domains and a cross-domain sequential recommendation (CDSRec) unit which takes the GUR as input to achieve cross-domain collaboration for the next-item recommendation task. \blue{Here, the GUR represents the static preference contains the user information in both domains.}

\red{The overall architecture of the proposed model is shown in Figure~\ref{fig:sys}. The GURU takes the long-term behavior sequence as input and \blue{generates generalized representations} for all users from both domains. Then, }the short-term user behavior together with the extracted general representation is fed into the CDSRec module to enhance sequential recommendation in each individual domain. 
\blue{To construct a GUR, we propose an adversarial training framework that involves an autoencoder for generating informative user representations and a domain discriminator to unify the representations across different domains. }
Furthermore, we apply an additional $l_2$ regularizer on overlapped users to further ensure each overlapped user has similar representations in both domains. 

\blue{Note that, motivated by the idea of the cross-lingual language models~\cite{lample2019cross}, we adopt a shared encoder across domains to avoid over-parameterization and achieve further information sharing via parameter-sharing across domains.}

Figure~\ref{fig:sys} gives a further example of how input sequences are processed to produce GURs and next-item recommendations. Specifically, the behavior sequences \blue{$(v_1^i,v_2^i,v_3^i , v_4^i,  [eos])$ and $(v_1^j, v_2^j, v_3^j, [eos])$ of user $u_i \in U_A$ and $u_j \in U_B$, where $[eos]$ denotes the end of sequence token}, 
are fed into two individual embedding layers to produce the embedding vectors \blue{$(\bm{e}_1^i, \bm{e}_2^i, \bm{e}_3^i, \bm{e}_4^i, \bm{e}_a^i)$ and $(\bm{e}_1^j, \bm{e}_2^j, \bm{e}_3^j, \bm{e}_b^j)$} containing both item information and sequence position information. 
Then, \red{the GURU encoder takes the} embedded sequences as inputs and generates the latent user representations $h_a^i$ and $h_b^j$ of users $u_i$ and $u_j$. To make sure the user representation is meaningful and informative, decoders are applied to reconstruct the original input sequences, such that, the encoder and decoder in each domain form an autoencoder framework. 
Furthermore, in order to get generalized representations which combines information from both the source and target domains, the domain discriminator is applied to $h_a^i$ and $h_b^j$ to pull the distributions of user representation in the two domains close to each other. Hereafter, in each domain, the sequential recommendation model combines the GUR and the short-term user behaviors to generate the next-item recommendation, \red{for example, the predicted item embeddings $\hat{\bm{I}}_4^j$ at the 4th timestamp of user $u_j$ in domain B.}

\subsection{Generalized User Representations}
As shown in Figure~\ref{fig:sys}, in the GURU module, user representations are learned in each individual domain, and information from the other domain is incorporated to make the representation ``general'' through regularizer achieved by domain discriminator on latent user representation. 

\textbf{User Representations in Single Domain}.
In each individual domain, we use an autoencoder to learn the latent user representation that is capable of reconstructing the original input sequence of the user. As compared to extracting user representations from the next-item prediction task, 
the autoencoder can produce meaningful representations through the reconstruction task 
to boost performance. 
The autoencoder in our framework consists of an embedding module, an encoder module, and a decoder module.

Formally, to extract a representation for any given user $u_i$ with its behavior sequence $s^i = (v_1^i, \cdots, v_t^i, \cdots, v_{|s^i|}^i)$, we apply the following procedures: 
\begin{equation} \label{eq:seq2seq}
\begin{aligned} 
\bm{e}_1^i, \cdots, \bm{e}_t^i, \cdots, \bm{e}_{|s^i|}^i, \bm{e}^i &= Embed(v_1^i, \cdots, v_t^i, \cdots, v_{|s^i|}^i, [eos]), \\
\bm{h}_1^i, \cdots, \bm{h}_t^i, \cdots, \bm{h}_{|s^i|}^i, \bm{h}^i &= Encoder(\bm{e}_1^i, \cdots, \bm{e}_t^i, \cdots, \bm{e}_{|s^i|}^i, \bm{e}^i), \\
v_1^i, \cdots, v_t^i, \cdots, v_{|s^i|}^i &= Decoder(\bm{h^i}),
\end{aligned}
\end{equation}
where $\bm{h}_t^i$ is the latent representation of item $v_t^i$ at position $t$, and $\bm{e}_t^i$ represents the sum of item embedding and positional embedding. 
$\bm{e^i}$ and $\bm{h^i}$ are the embedding and latent representation of the $[eos]$ token. We use bold font to indicate vector variables.

The input sequence of items and the $[eos]$ token are converted into real-valued vectors through the \textit{Embed} module, which consists of an item embedding layer and a positional embedding layer to incorporate both the item information and the sequential information of behavior sequences. 
To this end, we create two trainable embedding matrices $\bf{I} \subset \mathbb{R} ^{(|V| + 1)\times d}$ and $\bf{P} \subset \mathbb{R}^{(N + 1) \times d}$ for item and positional embeddings, respectively, where $d$ represents the number of dimensions in the latent space and $N + 1$ is the maximum length of input sequences including the $[eos]$ token. 
By summing up the output of item embeddings and positional embeddings with point-wise summation, we derive the embedding representations for all items and the $[eos]$ token, denoted as $(\bm{e}_1^i, \cdots, \bm{e}_{t}^i, \cdots, \bm{e}_{|s^i|}^i, \bm{e}^i)$ shown in Equation~\eqref{eq:seq2seq}. 

After the derivation of embeddings, we adopt the autoencoder to obtain the fixed-length representations of the behavior sequences of users. 
Specifically, we adopt an encoder with a structure similar to \emph{Transformer} \cite{vaswani2017attention} which is composed of a stack of identical transformer layers. Each transformer layer is composed of a multi-head, bidirectional self-attention layer, and a position-wise fully connected feed-forward layer.
\blue{Shown in Equation~\eqref{eq:seq2seq}, 
the \textit{Encoder} outputs a list of latent vectors $(\bm{h}_1^i, \cdots, \bm{h}_{|s^i|}^i, \bm{h}^i)$ for all input items in the input sequence.}
Then, the latent vector of the $[eos]$ token $\bm{h}^i$ is treated as the representation of the whole behavior sequence, i.e. user representation of $u_i$, and is further fed into the decoder. 

We adopt a decoder that also consists of multiple transformer layers. In addition to the self-attention and feed-forward sub-layers in each transformer layer, another multi-head attention layer over the user representation $\bm{h^i}$ is inserted right after the self-attention layer in each decoder layer. 
Additionally, the masking technique is applied to ensure that the predictions depend only on previous behavior sequences. 
Formally, shown in Equation~\eqref{eq:seq2seq}, our \textit{Decoder} takes the latent user representation $\bm{h}^i$ as input and reconstructs the original input sequence in an auto-regressive manner \cite{vaswani2017attention}.  

Illustrated in Figure~\ref{fig:sys}, different users have different lengths of behavior sequences, thus, we transform the input sequences of all users into fixed-length sequences with a length of $N + 1$, including the $[eos]$ token. 
Specifically, for users with sequences longer than $N$ (except for the $[eos]$ token), we consider the most recent $N$ items, for users with sequences shorter than $N$ we add $[pad]$ to the left of its original sequence until it has a total length of $N$. 

Following the convention of autoencoder, by optimizing the reconstruction error between the input samples and the reconstructed samples, our model can learn the most important attributes of the input behavior sequence. 
In this paper, the autoencoder structure is applied to sequence samples, thus we formulate the reconstruction loss for input sequence $s^i$ as follows:
\begin{equation} \label{eq:recons}
\begin{aligned}
\mathcal{L}_{rec}(s^i, \hat{s}^i) &= -\log p_{\theta}(\hat{s}^i|s^i)
\\
								&= -\sum_{t=1}^{N} \log p_{\theta}(\hat{v}_t^i | \hat{v}_{<t}^i, \bm{h}^i),
\end{aligned}
\end{equation} 	
where $s^i$ and $\hat{s}^i$ are the input and reconstruction output of autoencoder, and $\bm{h}^i = Encoder(Embed(s^i))$ denotes the latent user representation of user $u_i$ as is shown in Equation~\eqref{eq:seq2seq}. 
Moreover, at any position $t$, if its original item is the \textit{[pad]} token, we simply ignore the reconstruction loss at this position. 

\textbf{Generalizing User Representation Across Domains}. 
We propose to incorporate information from other domains into the user representation learned in a single domain to achieve cross-domain collaboration. 
Compared to the single-domain sequential recommendation, the representations regularized by the distributional constraints are more general and contain extra information from other domains.
A commonly used technique for autoencoder to integrate prior knowledge is to add extra penalty terms onto the reconstruction objective function:
\begin{equation} \label{eq:ae_constraint}
\mathcal{L}_{rec}(s^i, \hat{s}^i) + \Omega(\bm{h}^i),
\end{equation}
where $\Omega(\cdot)$ denotes the penalty function. 
For examples, sparse autoencoder (SAE) \cite{makhzani2013k} encourages sparsity of latent vectors by adding an $l_1$ or KL divergence penalty, variational autoencoder (VAE) \cite{kingma2013auto} assume that the latent representation $h^i$ follows a Multidimensional Gaussian distribution.

Similarly, to integrate knowledge from both domains, a Kullback–Leibler (KL) divergence penalty can be added to the reconstruction loss, where the KL divergence measures the distance between the distributions of the learned user representations in both domains,
leading to the following reconstruction objectives in both domains:
\begin{equation} \label{eq:cross_cons}
\begin{aligned}
& \mathcal{L}_{rec}(s_A, \hat{s}_A) + \textit{KL}(\rho^A||\rho^B), 
\\
& \mathcal{L}_{rec}(s_B, \hat{s}_B) + \textit{KL}(\rho^B||\rho^A),
\end{aligned}
\end{equation}
where $\rho^A$ and $\rho^B$ denote the learned distributions of latent user representations in domain $A$ and domain $B$, respectively. However, there is no closed-form expression for the distributions of latent user representations. Furthermore, the latent distributions are characterized by the parameters of the autoencoder, which is being constantly updated along the training process. 

Instead of directly estimating the ground-truth distributions, which is extremely difficult, we propose to bypass this challenge by adopting an adversarial training strategy to implicitly minimize the KL divergence between the two learned distributions $\rho^A$ and $\rho^B$.
As illustrated in Figure~\ref{fig:sys}, the encoders generate latent representations $\bm{h}_a^i \sim \rho^a$ and $\bm{h}_b^j \sim \rho^b$, where the distributions are characterized by the parameters of encoders in domain $A$ and $B$, respectively. 
A discriminator is then built for a binary classification task, where the input is a single latent representation either from domain $A$ or domain $B$, while the output is the prediction on which domain the input representation originates from. We adopt the neg log-likelihood loss as the objective function for the adversarial optimization process, denoted as:
\begin{equation} \label{eq:discriminator}
\begin{aligned}
\mathcal{L}_{dis}(h^i) &= -y \cdot log \sigma(f(h^i)) - (1-y) \cdot log(1-\sigma(f(h^i))), \\
y &= \mathbb{I}(u^i \in U^a)
\end{aligned}
\end{equation}
where $\mathbb{I}(\cdot)$ is the indicator function which equals to $1$ when the condition  $u^i \in U^a$ is true, otherwise, $y = 0$. $\sigma(\cdot)$ is the sigmoid function and the logits value $f(h^i)$ is calculated by the domain discriminator $f(\cdot)$.

Following the conventional optimization scheme of GAN, we alternately update the domain discriminator and the encoder until the model reaches the point where the discriminator is unable to tell whether a given user representation is from domain $A$ or $B$. 
At this point, the distributions of user representations $\rho^A$ and $\rho^B$ are supposed to be close to each other, i.e., which contributes to reducing the KL divergence $KL(\rho^A||\rho^B)$ between the two distributions. 
Therefore, we have achieved information sharing through the distributional constraint on latent user representations without relying on the information of overlapped users.

In addition to implicitly unifying the user representations with all users, we also propose an $l_2$ penalty to enforce the explicit representation sharing over domains for each overlapped user. That is, the representations of the same user should be the same or close across domains.
Illustrated in Figure~\ref{fig:sys}, for an overlapped user, i.e. $u_i=u_j$, we add an $l_2$ regularizer on its latent representations in domain $A$ and $B$:
\begin{equation} \label{eq:equal}
\mathcal{L}_{l_2} = ||\bm{h}_a^i - \bm{h}_b^j||_2, i=j.
\end{equation}
By adding the $l_2$ regularizer to the loss function for overlapped users, representations of the same user in different domains are pushed to be equal or close to each other, exploiting the information of overlapped users in an explicit way.


\SetKwInput{KwInput}{Input}
\SetKwInput{KwOutput}{Output}
\SetKwInput{KwMetaTraining}{Pre-training Phase}
\SetKwInput{KwInitialization}{Initialization}
\SetKwInput{KwTune}{Fine-tune}
\SetKwInput{KwPersonalization}{Multi-task Adversarial Training Phase}

\begin{algorithm}[tb]
\DontPrintSemicolon
\caption{RecGURU Training Algorithm}
\label{alg:gur_train}
\KwInput{
User sets $U^A$, $U^B$, item sets $V^A$, $V^B$; \\
Behavior sets: $S_A = \{s_A^i\}, \forall u_i \in U^A$, $S_B = \{s_B^j\}, \forall u_j \in U^B$; \\
Overlapped users $U^o = U^A \cap U^B$. \\
Hyper-parameter: CRITIC\_ITERS. \\
}
\KwInitialization{}
Initiate the item and positional embedding matrices: $\bf{I}^A \subset \mathbb{R} ^{(|V^A| + 1)\times d}$, $\bf{P}^A \subset \mathbb{R}^{ (N + 1) \times d}$ and $\bf{I}^B \subset \mathbb{R} ^{(|V^B| + 1)\times d}$, $\bf{P}^A \subset \mathbb{R}^{ (N + 1 )\times d}$. \\
Initiate the GURU, CDSRec and the domain Discriminator.\\ 
\KwMetaTraining{}

\red{Train autoencoders according to ~\eqref{eq:recons} on $S_A$ and $S_B$. }

\KwPersonalization{}
\While{not converged}
{
	\For{ $i\gets0$ \KwTo CRITIC\_ITERS }{
    	 Sample two batches of users $\bar{U}^A \sim U^A$, $\bar{U}^B \sim U^B$. 
         
         Optimize the discriminator loss~\eqref{eq:discriminator}.

    }

    Sample two batches of users $\bar{U}^A \sim U^A$, $\bar{U}^B \sim U^B$ and one batch of overlapped users $\bar{U}^o \sim U^o$.

    Fix the parameters of the domain Discriminator. 
    
    Train the model according to the loss defined in~\eqref{eq:loss_all}.

}

\KwTune{}
\red{Fine-tune the \emph{CDSRec} model according to the BPR loss defined in~\eqref{eq:loss_rc} in each domain.}
\end{algorithm}

\subsection{Cross-domain Sequential Recommendation}
The generalized user representation extracted by the proposed GURU module represents the overall preference of users in different domains, which is beneficial to the recommendation task in a specific domain. 

For the next-item recommendation task \red{in each domain}, a sequential recommender is built on the derived GURs and it is composed of multiple unidirectional attention layers and feed-forward layers. Specifically, the GURs are passed into the model through multi-head attention mechanisms. 

Formally, for a given user $u_i$ with sequence $s^i = (v_1^i, \cdots, v_{|s^i|}^i)$, the sequential recommender takes its generalized user representation $\bm{h}^i$ 
and short-term behaviors $(v_{|s^i|-m}^i, \cdots, v_{|s^i|}^i)$ as the input and outputs the current preference vector of the user $\bm{q}^{i, |s^i|}$ at time step $|s^i|$ in the latent space, given by
\begin{equation} \label{eq:final_rep}
\begin{aligned}
\bm{h}^i &= Encoder(Embed(v_1^i, \cdots, v_{|s^i|}^i, [eos])),
\\
\bm{q}^{i, |s^i|} &= CDSRec(\bm{h}^i, (v_{|s^i|-m}^i, \cdots, v_{|s^i|}^i)),
\end{aligned}
\end{equation}
where $m$ denotes the length of the short-term behavior sequence and the GUR is extracted from the long-term behavior sequence $(v_1^i, \cdots, v_{|s^i|}^i)$ of user $u_i$. \emph{CDSRec} denotes the cross-domain sequential recommender. 

The preference scores of the user to all the candidate items are then computed as the inner product between its current preference $\bm{q}^{i, |s^i|}$ and the item embeddings of all candidates, denoted as 
\begin{equation} \label{eq:final_prefer}
r^{i, |s^i|, v} = \bm{q}^{i, |s^i|} \red{\bm{I}_v}, \forall v \in V^c, \forall u_i \in U,
\end{equation}
where $\bm{I}_v$ is the item embedding of $v$ and the candidate set $V^c \subset V$ is a subset of the the entire item set $V$. 
Candidate items are then ranked and recommended according to the calculated preference scores. 

We adopt a Bayesian Personalized Ranking (BPR) loss \cite{rendle2012bpr} to train the recommendation model. 
For a given user $u_i$ from domain $A$, we calculate the loss of an item recommendation at time step $t$ as 
\begin{equation} \label{eq:loss_rc}
\mathcal{L}_{bpr} ^{t} = - \log\sigma(\bm{q}_A^{i, t} \bm{I}_v) - \log(1-\sigma(\frac{1}{|N_s|}\sum_{v' \in N_s}{} \bm{q}_A^{i, t} \bm{I}_{v'})),
\end{equation}
where $v$ and $\bm{I}_v$ are the target item and its corresponding item embedding, $N_s$ is the set of negative item samples and $\sigma(\cdot)$ represents the sigmoid function. 

\subsection{Training Strategy} 
\label{subsec:model_train}

We propose a three-phase training algorithm shown in Algorithm~\ref{alg:gur_train} to optimize the proposed model. 

In the first phase, we pre-train the autoencoders in each domain individually with the reconstruction task. Through the pre-training process, the reconstruction loss is largely reduced, producing a boost-start for the following adversarial training. 
In the second phase, following the training process of GAN in~\cite{gulrajani2017improved}, at each iteration, we first optimize the discriminator loss $\mathcal{L}_{dis}$ for \textit{CRITIC\_ITERS} steps (which equals to 5 in our implementation). Then, the reconstruction task, negative discriminator loss, and $l_2$ loss on overlapped users are jointly optimized in a multi-task fashion by minimizing the loss: 
\begin{equation} \label{eq:loss_all}
\red{\mathcal{L} = \mathcal{L}_{rec} - \mathcal{L}_{dis} + \mathcal{L}_{l_2}.}
\end{equation}
With the reconstruction task, we prevent encoders from generating wild representation stabilizing the adversarial learning. 
Following the common practice in GAN training~\cite{gulrajani2017improved}, we also adopt the gradient penalty term in the critic optimization step. 
\red{Finally, we fine-tune the \emph{CDSRec} model in each individual domain with the next-item recommendation task. }

\section{Experiments}
\label{sec:exp}
In this section, we conduct extensive experiments on two public and one collected cross-domain sequential recommendation scenarios. 
Comparison with the state-of-the-art single domain and cross-domain baselines shows the effectiveness of our proposed method. Furthermore, ablation tests demonstrate the impact of each proposed sub-modules on the recommendation result and the robustness of our model under scenarios of a different portion of overlapped users.

\subsection{Dataset and Experiment Setup}
\label{subsec:exp_set}

\begin{table}[tbp]
\small
\caption{Statistics for the three cross-domain scenarios.
} 
\begin{center}
\begin{tabular}{ccccc}
\toprule
Domain & \#Users & \#Items & Avg. Seqlen. & \#Overlap. \\
\hline
Sport & 9,024  & 11,835 & 6.62   & \multirow{2}{*}{1,062}        \\        
Cloth & 46,810 & 42,139 & 7.51   &         \\        
\hline
Movie & 4,261  & 5,536  & 8.307 &  \multirow{2}{*}{584}           \\ 
Book  & 42,940 & 51,366 & 9.490 &           \\ 
\hline
\red{Wesee}    & 1,952,403  & 335,648 &  18.196  &     \multirow{2}{*}{1,692,893} \\      
\red{Tencent Video}  & 2,183,927  & 1,455,595 &  28.53   &     \\      
\bottomrule
\end{tabular}
\label{table:Dataset_info}
\end{center}
\end{table}

Four publicly available Amazon datasets \cite{ni2019justifying} and two collected datasets are used to form three cross-domain sequential recommendation datasets: ``Sports-Clothing'', ``Movie-Book'', and the collected dataset. 
On Amazon datasets, only recent positive reviews, posted after October 1, 2017, with a rating score higher than 2, are selected. 
We collect data from two popular video applications, ``Wesee'' and ``Tencent Video'', both with over billions of daily active users. 
Watching histories for three consecutive days,from June 26, 2020, to June 28, 2020, of these two applications are collected in this dataset. The detailed breakdown of the three cross-domain sequential recommendation datasets is shown in Table~\ref{table:Dataset_info}. 
Note that, the collected dataset mostly consists of overlapped users, thus, we can manually adjust the portion of overlapped users in a wide range to test the robustness of our method.

Following common practices in sequential recommendation~\cite{sun2019bert4rec}, for a given user the second to last item in the behavior sequence is selected as the validation item, and the last item is used for testing, while the remaining items are used for training. 

\begin{table*}[tbp]
\small
\begin{center}
\caption{Comparison between the proposed method with single-domain and cross-domain baselines on ``Sport-Cloth'' and ``Movie-Book'' scenarios. \red{\textbf{``AutoRec''} and \textbf{``AutoEM''} are two variants of our proposed method. All values are in percentage}.} 
\begin{tabular}{ll|ccccc|cccccc}
\toprule
\multirow{2}{*}{Datasets} & \multirow{2}{*}{Metric}  &  \multicolumn{5}{c|}{Single-domain algorithms} &   \multicolumn{6}{c}{Cross-domain algorithms} \\
\cline{3-13}
& & POP & BPRMF & SAS & Bert4Rec & \textbf{AutoRec} & CMF & MFEM & CnGAN & \blue{DOML} &\textbf{AutoEM} & \textbf{RecGURU} \\ 
\midrule
\multirow{4}{*}{Sport} 
& HR@5    & 3.63  & 16.17 & 17.51    & 18.04 & 18.82   & 17.04 & 15.87 & 15.88 & \red{15.26} &17.63   & \textbf{20.78}\\
& HR@10   & 5.70  & 19.62 & 22.09    & 22.46 & 23.37   & 20.55 & 19.44 & 19.41 & 20.29 & 22.57  & \textbf{24.88}\\
 & NDCG@5  & 3.03  & 12.43 & 14.47    & 14.57 & 14.93  & 13.37 & 12.68 & 12.89 & 11.66 & 14.06  & \textbf{16.56}\\
 & NDCG@10 & 3.70  & 13.73 & 15.94    & 16.01 & 16.36  & 14.45 & 13.77 & 14.05 & 13.17 & 15.61  & \textbf{17.87}\\
\cline{2-13}
\multirow{4}{*}{Cloth} 
& HR@5    & 1.48  & 21.03 & 23.71    & 22.68 & 23.89   & 21.53 & 20.56 & 20.10 & 22.70 & 24.27  & \textbf{25.68}        \\
& HR@10   & 2.10  & 23.98 & 28.40    & 26.57 & 27.67   & 24.33 & 23.41 & 23.29 & 28.36 & 28.71  & \textbf{29.16} \\
& NDCG@5  & 0.08  & 17.89 & 19.81    & 19.95 & 20.51   & 18.44 & 17.51 & 16.72 & 17.47 & 20.44  & \textbf{22.44} \\
& NDCG@10 & 0.96  & 18.86 & 21.29    & 21.20 & 21.70   & 19.30 & 18.40 & 17.70 & 19.29 & 21.83  & \textbf{23.52} \\
\hline                       
\multirow{4}{*}{Movie} 
& HR@5    & 2.07  & 12.80 & 12.18    & 13.78 & 14.21   & 12.34 & 14.70 & 14.03 & \red{14.41} & 12.93  & \textbf{18.97}    \\
& HR@10   & 5.37  & 17.28 & 17.20    & 19.39 & 19.51   & 15.46 & 19.52 & 18.49 & 19.78 & 18.94  & \textbf{24.24} \\
& NDCG@5  & 1.26  & 9.36  & 8.76     & 9.79  & 10.15   & 8.36  & 10.80 & 10.18 & 9.75 & 8.67   & \textbf{14.59} \\
& NDCG@10 & 2.30  & 10.85 & 10.38    & 11.55 & 11.83   & 8.75  & 12.31 & 11.57 & 11.94 & 10.61  & \textbf{16.17} \\
\cline{2-13}                   
\multirow{4}{*}{Book}  
& HR@5    & 2.69  & 22.37 & 27.15    & 27.86 & 28.10   & 22.70 & 23.17 & 23.24 & 20.67 &  \textbf{28.92} & 28.15 \\
& HR@10   & 5.31  & 31.28 & 36.85    & 37.58 & 38.20   & 31.56 & 32.33 & 32.33 & 30.86 & \textbf{38.63} & 38.35 \\
& NDCG@5  & 1.79  & 16.32 & 19.50    & 19.93 & 20.09   & 16.38 & 16.63 & 16.63 & 14.18 & \textbf{20.89} & 20.07 \\
& NDCG@10 & 2.62  & 19.09 & 22.60    & 23.03 & 23.33   & 19.19 & 19.36 & 19.36 & 17.14 & \textbf{23.99} & 23.06 \\
\bottomrule
\end{tabular}
\label{table:result_pub}
\end{center}
\end{table*}

Hit Ratio (HR) and Normalized Discounted Cumulative Gain (NDCG) \cite{jarvelin2002cumulated} are adopted to evaluate the performance of all methods. 
We follow the strategy used in \cite{he2017neural} to reduce the heavy computation cost. Specifically, for users from public datasets, we sample 200 negative items in the item list with respect to their frequencies, which together with the ground-truth item, form the candidates for recommendation. On collected datasets, the size of the candidate set becomes 20,000. HR and NDCG with $k = 5, 10, 20$ are reported. The following single-domain and cross-domain baseline algorithms are evaluated. 
\begin{itemize}[leftmargin=*]
\item \textbf{POP}: All items are recommended according to their popularities. 
\item \textbf{BPRMF}~\cite{rendle2012bpr}: It optimizes the matrix factorization with the BPR loss.
\item \textbf{SAS}~\cite{kang2018self}: \red{It adopts} unidirectional self-attention to model user behaviors.
\item \textbf{Bert4Rec} \cite{sun2019bert4rec}: It incorporates the idea of Bert~\cite{devlin2018bert} to the next item recommendation task.
\item \textbf{AutoRec}: It is the single-domain version of proposed model that adopts autoencoder to generate static user representation which is used for the sequential recommendation. 
\item \textbf{CMF}~\cite{singh2008relational}: It simultaneously factors interaction matrices in both target and source domains. 
\item \textbf{MFEM}: It learns a mapping function on overlapped users. Thus, for non-overlapped users, we can get their embeddings in the source domain through the trained mapping function as well as the cross-domain recommendation through Equation~\eqref{eq:cross_seq}. Here, the user embeddings are learned with the BPRMF model. 
\item \textbf{CnGAN}~\cite{perera2019cngan}: \red{It adopts adversarial learning to learn a better mapping function that maps user embeddings from the source domain to the target domain or vice versa. }
\item \textbf{DOML} \cite{li2021dual}: \blue{It adopts dual metrics learning in cross-domain recommendation when there are few overlapped users.}
\item \textbf{AutoEM}: As a variation of proposed method, it also learns a mapping function on overlapped users to get the embeddings of non-overlapped users. However, here, the item and user embeddings are learned by the proposed AutoRec model. 
\end{itemize}

\subsection{Implementation Details}
We implement the models using PyTorch with python 3.6 and train our framework on Tesla P40 GPUs with a memory size of 22.38 GiB and a 1.53 GHz memory clock rate.

\blue{On Amazon datasets, 3 transformer layers are adopted, whereas 6 transformer layers are used on the collected datasets in the encoder and decoder module. We use 2 attention heads for all the attention layers throughout the model. And the dimension of all feed-forward layers is set to 512. }
For adversarial training, we build the domain discriminator with four fully connected layers with a hidden size of 128. Furthermore, we adopt the improved W-GAN \cite{gulrajani2017improved} framework to alternately optimize the domain discriminator and the GURU model through the discriminator loss $\mathcal{L}_{dis}$. 
For simplification, we adopt the same length of sequence for both autoencoder and recommendation tasks, which is 100 on all datasets. The number of dimensionality of user embedding is 64 on both public and collected datasets. 
Multiple \blue{Adam optimizers \cite{kingma2014adam} are used to update different modules of the proposed RecGURU framework.}

For all the transfer learning-based cross-domain baselines, we first concatenate the user embeddings from the source and target domains. Then, a fully connected layer is applied to get the cross-domain user embedding which is further used for next item recommendation. 
More details are given in the supplementary.

\subsection{Experimental Results and Analysis}
\begin{table*}[tb]
\small
\caption{Ablation studies on customized collected datasets with the portion of overlapped users ranging from 10\% to 75\%. \red{All values are in percentage.}} 
\begin{center}
\begin{tabular}{llcccc|cccc|cccc}
\toprule

\multicolumn{2}{c}{Variants methods} & \multicolumn{4}{c}{SeqRec} & \multicolumn{4}{c}{\textbf{+Auto}}  & \multicolumn{4}{c}{\textbf{+GURU}}                                                                     \\ 
\hline
Dataset & Metric  & 10\%  & 30\%  & 50\%  & 75\%  & 10\% & 30\%  & 50\%  & 75\%  & 10\% & 30\% & 50\% & 75\% \\ 
\midrule
 \multirow{4}{*}{Wesee}    & HR@5     & 14.58 & 15.19 & 15.19 & 15.59 & 15.41          & 15.49 & 15.52 & 16.12 & \textbf{16.13}               & \textbf{17.11}               & \textbf{17.15} & \textbf{16.59} \\
 & HR@10    & 21.72 & 22.47 & 22.39 & 23.08 & 22.71          & 22.84 & 23.08 & 23.9  & \textbf{23.61}               & \textbf{24.87}               & \textbf{24.87} & \textbf{24.34} \\
 & NDCG@5   & 9.67  & 10.1  & 10.05 & 10.31 & 10.36          & 10.55 & 10.37 & 10.89 & \textbf{10.91}               & \textbf{11.64}               & \textbf{11.66} & \textbf{11.31} \\
 & NDCG@10  & 11.94 & 12.43 & 12.34 & 12.68 & 12.71          & 12.91 & 12.8  & 13.39 & \textbf{13.34}               & \textbf{14.14}               & \textbf{14.11} & \textbf{13.79} \\
\cline{2-14} 
 \multirow{4}{*}{Tencent Video}  & HR@5     & 25.33 & 25.95 & 27.14 & 26.71 & \textbf{28.48} & \textbf{29.21} & 31.99 & 31.60 & 28.26                        & 29.14      & \textbf{32.18} &  \textbf{31.23} \\
 & HR@10    & 32.56 & 33.88 & 35.89 & 34.92 & 35.8           & 36.41 & \textbf{39.76} & \textbf{39.76} & \textbf{36.11}               & \textbf{37.12} &38.49  &  39.14  \\
 & NDCG@5   & 18.62 & 18.66 & 19.09 & 18.91 & \textbf{21.49}          & \textbf{21.7}  & \textbf{23.91} & \textbf{24.33} & \red{20.78} & \red{21.41} & 23.12&  23.39\\
 & NDCG@10  & 20.95 & 21.16 & 21.91 & 21.55 & \textbf{23.87}          & \textbf{24.01} & \textbf{26.4}  & \textbf{26.78} & \red{23.32} & \red{23.98} &25.51 &  25.95\\
\bottomrule
\end{tabular}
\label{table:result_ab}
\end{center}
\end{table*}

\begin{figure}[tbp]
\small
    \centerline{
        \subfigure[losses of the adversarial training]{        
            \includegraphics[width=1.5in]{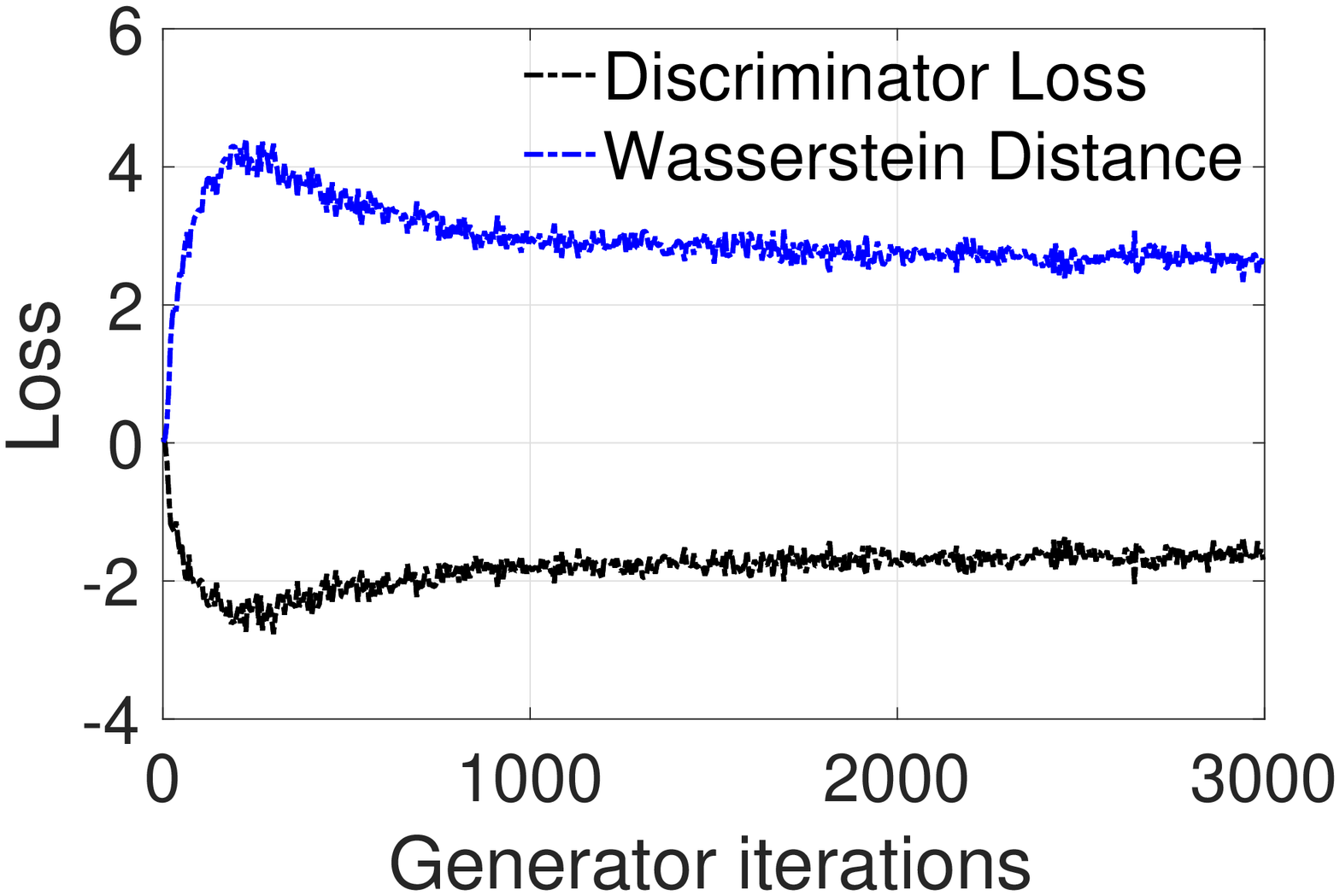}
            \label{fig:loss_gan}
        }
        \hspace{0.1in}
        \subfigure[Reconstruction and BPR losses]{
            \includegraphics[width=1.5in]{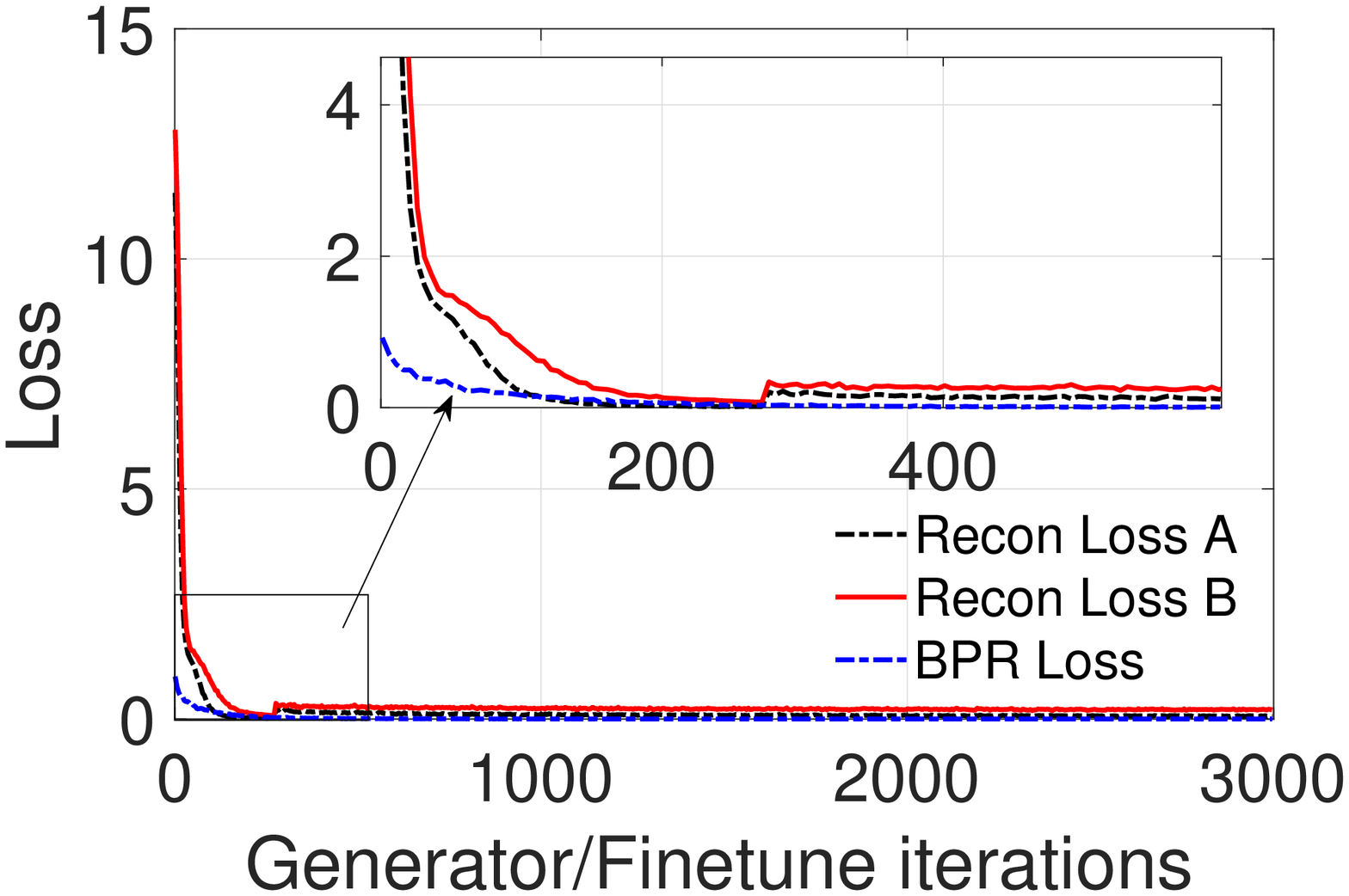}
            \label{fig:loss_rec}
        }
    }
    \caption{Training losses on the ``Movie-Book'' dataset. }
    \label{fig:training_loss}
    \Description{loss}
\end{figure}

\textbf{\red{Training Loss. }}
Figure~\ref{fig:training_loss} shows the training losses on the ``Movie-Book'' dataset. Specifically, the Wasserstein distance and the critic loss, i.e. discriminator loss, are given in Figure~\ref{fig:loss_gan}. Both processes converge around a thousand adversarial iterations. In the beginning, the discriminator is good at distinguishing representations from different domains with an increasing Wasserstein distance. After more adversarial training iterations, the Wasserstein distance is reduced and converged which in line with the training process of standard GANs~\cite{gulrajani2017improved}. 
The reconstruction tasks in both target and source domains converge after 400 iterations as is shown in Figure~\ref{fig:loss_rec}. The recommendation task in the target domain converges after only 100 steps of fine-tune iterations.

\textbf{\red{Main Results.}}
Table~\ref{table:result_pub} summarizes the performances of all baselines and the proposed methods on the two public cross-domain scenarios. 
AutoRec, the single domain version of our solution, outperforms SAS and Bert4rec on single domain sequential recommendation tasks on most of the public datasets. 
The success of AutoRec can be attributed to the adoption of user representations extracted from the history user behavior through an autoencoder.

The proposed RecGURU outperforms all other baselines with an improvement on HR@10 and NDCG@10 of \blue{$6.5\%$, $9.2\%$, and $ 22.5\%$, $31.3\%$} on the ``Sports'' and ``Movie'' datasets, respectively, compared to the best given by all the other baselines. 
We can also outperform the state-of-the-art single-domain and cross-domain methods on ``Cloth'' and ``Book'' datasets in most cases but with a relatively smaller margin compared to the results reported on the ``Sport'' and ``Movie'' datasets. This is reasonable, as is shown in Table~\ref{table:Dataset_info}, the ``Sport'' and ``Movie'' datasets are much smaller and with more sparsity than the ``Cloth'' and ``Book'' datasets. Obviously, when the data in the source domain is highly sparse, we can only get a limited amount of information from the source domain to help with recommendations in the target domain. 
\red{Therefore, AutoRec, AutoEM, and RecGURU achieve close performance on the ``Book'' dataset. However, overall RecGURU outperforms other baselines in most cases. }
The superiority of RecGURU over AutoRec and AutoEM can be attributed to the generalization of user representations achieved through the adversarial training of the domain discriminator and the autoencoder.
\red{Furthermore, compared with the improvements RecGURU has achieved over single domain sequential recommendation baselines, cross-domain methods such as MFEM, CnGAN, and DOML achieved limited improvements over BPRMF. This can be attributed to the fact that these baselines either rely on overlapped users (MFEM, CnGAN) or need side information for more general user and item embeddings (DOML).} 

\subsection{Ablation Study}
Variants of the proposed method are evaluated on the collected datasets for ablation studies to show the impact of each proposed sub-modules. We incrementally accommodate different modules into the single-domain sequential recommendation model SeqRec, until we incorporate all the proposed sub-modules and features. Specifically, the following models are evaluated:
\begin{itemize} 
    \item \textbf{SeqRec}: Sequential recommendation model without autoencoder.
    \item \textbf{+Auto}: The AutoRec model introduced in Sec.~\ref{subsec:exp_set}.
    \item \textbf{\red{+GURU}}: The proposed RecGURU model. 
\end{itemize}
Furthermore, to test the robustness of the proposed method under a variety of overlapping rates, we manually change the number of overlapped users to form four new datasets with an overlapping rate of 10\%, 30\%, 50\%, and 75\%. 

Table ~\ref{table:result_ab} summarizes the results of ablation tests of all the introduced variants on all datasets with various overlapping rates. 
On the ``Wesee'' dataset, each time we add a new sub-module or feature incrementally on top of the previous model, we can observe improvement in the overall recommendation performance, which illustrates the effectiveness of autoencoder and GURU modules. 
Moreover, this phenomenon appears on all the four datasets with overlapping ranging from 10\% to 75\% which shows the robustness of our method to user overlapping rate. 
Similar to the ``Book'' dataset, on the ``Tencent Video'' dataset, the single-domain version of our proposed method, AutoRec, is slightly better than its cross-domain version, which is also due to the sparsity issue in the source domain, i.e. ``Wesee'', as we have explained before.
 
 \section{Conclusion}
\label{sec:conclusion}
In this paper, we propose RecGURU, a novel cross-domain sequential recommendation framework based on Generalized User Representations (GURs). Different from previous work which aims to transfer knowledge across domains, in the RecGURU system, we propose the GURU module that is capable of extracting a generalized user representation unified over different domains via adversarial learning. 
Specifically, an autoencoder is adopted to generate user representations in each individual domain. Cross-domain generalization of user representations is achieved by adversarially training a discriminator and the encoder until the domain-dependent embeddings are statistically indistinguishable among different domains. We further propose various schemes to stabilize and boost the learning effectiveness of RecGURU. 
Experimental results on both publicly available datasets and collected datasets show the effectiveness of the proposed method.

\bibliographystyle{ACM-Reference-Format}


\newpage

\section{SUPPLEMENTARY}

\subsection{Model Implementation details}
As mentioned in the paper, the proposed RecGURU is composed of two parts: the generalized user representation unit (GURU) and the cross-domain sequential recommendation (CDSRec) module. Here we introduce the detailed implementation of these two modules.

\subsubsection{GURU}.The generalized user representation unit (GURU) consists of one shared encoder and two decoders which form two Autoencoder frameworks in both the target and source domains. Furthermore, a domain discriminator is adopted to unify user representations from both domains in an adversarial learning scheme.

\textbf{Encoder}:
As we have mentioned in the paper, the encoder is composed of several transformer layers where each transformer layer consists of one multi-headed self-attention layer and a fully connected feedforward layer. Specific hyper-parameters of the encoder are:
\begin{itemize} 
	\item \textit{Number of layers}: The number of transformer layers used to build the encoder. We have grid searched the number of transform layers in the range of $[2,3,4,5,6]$ on all public datasets and finally decided to adopt $3$ layers on all public datasets, i.e. the ``Movie'', ``Book'', ``Sport'' and ``Clothing'' datasets. On the collected datasets, i.e. ``Wesee'' and ``Tencent Video'', 6 layers of transformer layers are adopted to build the encoder. 
	\item \textit{Number of heads}: Number of heads for all multi-head attention layers within each transformer layer, which equals 2 in our model.
	\item \textit{Dimension of embedding}: The dimensionality of the latent space for user representation and item embedding, which is set to be 64. To have a fair comparison between different methods, the number of dimensions of the item and user embeddings are all set to be 64 for all experiments that appeared in this paper.
	\item \textit{Dimension of feedforward layer}: We adopt feed-forward layers of size 512 for all transformer layers.
\end{itemize}	

\textbf{Decoder}: we use the same hyperparameters as the encoder to build decoders in each domain (datasets), the only difference is that for each transformer layer in the decoder, we insert another attention layer to incorporate latent user representation generated by the encoder.

\textbf{Domain Discriminator}:
The domain discriminator, as a binary classification model, takes a user representation, from either the source domain or the target domain, as input and predicts the origin domain that the user comes from. In this paper, the domain discriminator is composed of several fully connected neural layers. Specifically, the hyper-parameters of the domain discriminator are defined as:
\begin{itemize}	
	\item \textit{Number of layers}: Four fully connected layers are used in our domain discriminator. 	
	\item \textit{Middle dimension}: The dimensionality of the middle layers in our domain discriminator, which is set to be double the size of the latent vector, i.e. 128.
\end{itemize}

\subsubsection{CDSRec}
We adopt the same model structure as the decoder for the cross-domain sequential recommendation. However, the input of the CDSRec is the short-time activity of a user which aims to model users' dynamic preference in the target domain. The long-term interest and cross information are provided by the GUR (generalized user representation) given by the GURU module. Through attention layers, we are able to combine the GUR with the user's short-term preference to provide a more precise recommendation in the target domain. 

\subsubsection{Model Training}
The training procedure of the proposed RecGURU has been introduced in the paper (Sec. 3.5), here we provide some important hyper-parameters and related to model training for reproducibility. Specifically, the following aspects are discussed here: 

\textbf{Optimizer}: 
The ADAM optimizer is used in our work to train all the modules. We adopt different ADAM optimizers with different learning rates at each training phase of the RecGURU Training Algorithm. Specifically, at the pre-training phase, we use a wrapped ADAM optimizer with a starting learning rate of $1.0$, $\beta_1 = 0.9, \beta_2=0.98$ and $ \epsilon = 1 \times 10^ {-8}$, the learning rate of the wrapped optimizer is adjusted from epoch to epoch. In the adversarial optimization procedure, two optimizers are adopted for both the critic step, which optimize the domain discriminator loss, and the generator step, which optimize the negative discriminator loss along with the reconstruction and $l_2$ constraint loss on overlapped users. We refer these two optimizers as \textit{opt\_dis} and \textit{opt\_gen}, respectively. For these two optimizers we adopt a learning rate of $ 1\times 10^{-4}$, $\beta_1 = 0.5$, $\beta_2=0.9$ and $ \epsilon = 1 \times 10^ {-8}$. Finally, in the fine-tune stage, another ADAM optimizer with an learning of $ 1\times 10^{-3}$, $\beta_1 = 0.9$, $\beta_2=0.9$ and $ \epsilon = 1 \times 10^ {-8}$ is adopted.

\textbf{Sampled Soft-max}: As defined in the paper, the reconstruction loss of the autoencoder is given by:
\begin{equation} \label{eq:recons_sup}
\begin{aligned}
\mathcal{L}_{rec}(s^i, \hat{s}^i) &= -\log p_{\theta}(\hat{s}^i|s^i) \\								&= -\sum_{t=1}^{N} \log p_{\theta}(\hat{v}_t^i | \hat{v}_{<t}^i, \bm{h}^i).
\end{aligned}
\end{equation} 	
Following previous studies, such as BERT, we use cross-entropy to model the probability distribution $p_{\theta}$ over all possible outputs (the entire item set) at each position of the reconstructed sequence. However, this may cause memory error when the number of items is huge, for example, we have hundreds of thousands of items in our collected datasets. To solve this problem, we adopted the sampled soft-max to avoid memory issues. Specifically, we first randomly select neg\_softmax negative samples and we perform softmax and calculate the probability distribution $p_{\theta}$ over the sampled negative samples and the target sample (the ground truth next-item). Finally, cross-entropy is calculated over the sampled items and the ground-truth item. This way, we can reduce the calculation burden and avoid memory error on the dataset with a large number of items.The number of chosen negative samples is set to be 20 on all the public datasets.On the collected dataset, it is set to be 10. 


\textbf{Gradient penalty}: We also adopted the gradient penalty to further stabilize the training of the adversarial networks. The number of critic steps in each adversarial iteration \textit{CRITIC\_ITERS} is set to be 5.

\subsection{Implementations of Baselines}
Most of the single domain baseline methods have open-sourced their implementation. For a fair comparison, for any shared hyper-parameters, for example, the dimensionality of the item and user embedding, the number of transformer layers for the SAS and Bert4Rec model, we set them to be the same as what we used in our model (please refer the previous section). For any other model-specific hyper-parameters, we use the default value provided in their opened source code. In summary, we have referred to the following open-sourced projects for all the baselines: 
\begin{itemize}[leftmargin=*]
	\item \textbf{BPRMF}: \url{https://github.com/guoyang9/BPR-pytorch}
	\item \textbf{SAS}: \url{https://github.com/kang205/SASRec}\item 
	\textbf{CMF}: Minor revision from the BPRMF code.
	\item \textbf{DOML}: \url{https://github.com/lpworld/DOML}
\end{itemize}
For those baselines that haven't open sourced their code, we either implemented it by ourselves, for example, MFEM is the combination of BPRMF and a mapping function of user embeddings from the source domain to the target domain, or request the authors for their code and adapt it to our problem, for example, we have implemented our version of CnGAN based on the requested code from the authors of CnGAN. The implementations of baselines are all included in our code package\footnote{\url{https://github.com/Chain123/RecGURU}}.

Note that, the origin DOML is designed for cross-domain rating prediction, however, in this paper, we focus on the implicit recommendation. Therefore, we made a minor modification on the origin code and the MSE with the BPR-loss. 


\subsection{Dataset and Pre-processing}
\subsubsection{The Amazon datasets}
We applied the following pre-processing steps on all four Amazon datasets. First, we selected all the ratings with a score larger than 2, thus, all the interactions in the selected dataset are supposed to be positive feedback. We further ignore the review scores for the implicit feedback-based recommendations. Second, we get the 5-core dataset based on the selected interactions. Third, we re-code all the items from 1 to the total number of items in each dataset. Finally, we extracted the user IDs for all the overlapped users in the two cross-domain scenarios, i.e. ``Sport-Cloth'' and ``Movie-Book''.

\subsubsection{The Collected dataset}\footnote{\url{https://drive.google.com/file/d/1NbP48emGPr80nL49oeDtPDR3R8YEfn4J/view?usp=sharing}}
To facilitate the recommendation in video streaming and short videos, we collect two sequential video recommendation datasets from two real-world applications, ``Wesee'' and ``Tencent Video'', which provide all kinds of short videos and dramas to hundreds of billions of users.``Wesee'' is a Chinese video streaming website providing popular movies \red{and all kinds of TV shows} as well as short videos to over one hundred million users. On the other hand, ``Tencent Video'' is a short video \red{sharing} platform with over ten million daily active users.

\vfill\eject
Note that, in the real scenario, there are only a few portions of overlapped users between these two applications, however, here we chose most of the overlapped users due to: first, there are still hundreds of thousands of overlapped users, second, with high overlapping rate, we are able to adjust the overlapping rate manually for stress tests. 

To form a cross-domain sequential recommendation scenario, the two datasets are collected within the same time period so that user preferences are consistent during dataset collection, and thus can be transferred between domains for cross-domain collaboration. For each application, interaction information such as action, item ID, and time-stamp is collected for all users.We consider interactions with ``click'', ``finish'' (which means finish watching the whole video) as positive interactions between users and video items, all other interactions are ignored. For a given user, its behavior in three consecutive days is collected and stored as three lists: action list, item ID list, and time-stamp list. Then we remove entries with \red{bad user IDs and item IDs, such as empty or garbled codes}. Furthermore, to avoid duplicated redundancies, if an item appears in a raw in the behavior sequence for the same user, we drop all the redundant repeats and only keep one copy. Finally, users with behavior sequences of length larger than 5 are used in our paper. Detailed statistics of these two datasets are given in our submission.

\begin{table}[tbp]
\small
\caption{Dataset statistics of the four collected cross-domain scenarios with the portion of overlapped users ranging from 10\% to 75\%. }
\begin{center}
\begin{tabular}{llcccc}
\toprule
Scenario              & Dataset & \multicolumn{1}{l}{\#User} & \multicolumn{1}{l}{\#Item} & \multicolumn{1}{l}{Avg.Seqlen} & \multicolumn{1}{l}{\#Overlap} \\
\hline
\multirow{2}{*}{10\%} & Wesee         & 1203194     & 288155    & 16.16 & \multirow{2}{*}{194756}       \\
                      & Tencent Video & 1434999     & 1311440   & 26.46 &                               \\
\hline                      
\multirow{2}{*}{30\%} & Wesee         & 1398645     & 303942    & 16.18 & \multirow{2}{*}{585580}       \\
                      & Tencent Video & 1630372     & 1356831   & 26.49 &                               \\
\hline                      
\multirow{2}{*}{50\%} & Wesee         & 1594256     & 316338    & 16.19 & \multirow{2}{*}{976556}       \\
                      & Tencent Video & 1825737     & 1397521   & 26.51 &                               \\
\hline                      
\multirow{2}{*}{75\%} & Wesee         & 1838276     & 329679    & 16.19 & \multirow{2}{*}{1464731}      \\
                      & Tencent Video & 2069892     & 1438803   & 26.53 &                              \\
\bottomrule
\end{tabular}
\label{table:Dataset_info_sup}
\end{center}
\end{table}

\textbf{Datasets with different portions of overlapped users.} To test the robustness of our method under different portions of user overlapping scenarios, we form four new datasets from the originally collected dataset by randomly sampling the overlapping users. Table~\ref{table:Dataset_info_sup} provides an overview of the four manufactured datasets.?As we did not include users real ID to protect their privacy,For each overlapped user we generate a random number to decide whether to keep it as an overlapped user or put it to a specific domain, i.e. ``Wesee'' or ``Tencent Video''. If an overlapped user is put to a specific domain as a ``non-overlapped'' user, its behavior sequence in another domain is removed. Therefore, in the different manufactured datasets, we would have a slightly different number of items.

\end{document}